\documentstyle[preprint,eqsecnum,aps,epsf,psfig]{revtex}
\newif\iftightenlines\tightenlinesfalse
\tightenlines\tightenlinestrue
\begin{document}

\title{\hfill { \bf {\small DO-TH 00/10}}\\[1cm] 
Single Pion Production in Neutrino Reactions and Estimates for Charge-Exchange Effects.}
\author{E. A. Paschos\footnote{On sabbatical from the University of Dortmund}}
\address{
Institute for Advanced Study, Princeton, NJ 08540,
U.S.A.}
\author{L. Pasquali and J. Y. Yu}
\address{
Institut f$\ddot{u}$r Physik, Universit$\ddot{a}$t Dortmund, D-44221 Dortmund,
Germany}

\maketitle

\begin{abstract}
We calculate single pion production by neutrinos in the resonance region. We consider both charged and neutral current reactions on free protons and neutrons. We present differential and total cross sections which can be compared with experiments. Then we use these results to calculate the spectra of the emerging pions including the Pauli suppression factor and rescattering corrections for reactions in heavy nuclei. Our results will be useful for studying single pion production and for investigating neutrino oscillations in future experiments.
\end{abstract}

\section{INTRODUCTION}

There is strong evidence for the mixing of muon neutrinos with another state, being either tau or sterile neutrinos. The evidence comes from atmospheric neutrino experiments which observe a decrease of muon neutrinos in charged current reactions, but no decrease of the corresponding electron neutrino interactions \cite{Fukuda}.

In order to obtain better insight into the oscillation which takes place and in order to eliminate dependence on the flux there are proposals and experiments being planned and constructed, which look at the neutral current interactions. These are reactions which will use neutrinos of an average energy of 1 GeV producing the resonances between 1.0 and 1.6 GeV/c$^2$. One proposal considers the production of pions directly by the atmospheric neutrinos \cite{Vissani} and the detection of $\pi^0$'s with the help of two ring events. This method is limited by the low flux of atmospheric neutrinos.

More powerful are experiments which use neutrinos from an accelerator with two detectors; the first one nearby the accelerator and a second further away. The nearby detector will be able to detect all pions and check the flux, as well as the cross-sections for these reactions. The detector with the long-baseline (300-400 km) will observe the charged and neutral current reactions.

We classify the reactions. Quasi-elastic scattering $\nu_\mu + n \rightarrow \mu^- + p$ is well understood. The oscillation of muon neutrinos into other neutrinos, tau or sterile, will produce a reduction of muon events in the far away detector.

In addition to the above reaction there are excitations of resonances and their subsequent decays
\begin{equation}
\nu_\mu + p \rightarrow \mu^- + p +\pi^+
\end{equation}
\begin{equation}
\nu_\mu + n \rightarrow \mu^- + n +\pi^+
\end{equation}
\begin{equation}
\nu_\mu + n \rightarrow \mu^- + p +\pi^0 \ .
\end{equation}
 
Furthermore, there are the neutral current reactions
\begin{equation}
\nu_\mu + p \rightarrow \nu_\mu + p +\pi^0
\end{equation}
\begin{equation}
\nu_\mu + p \rightarrow \nu_\mu + n +\pi^+
\end{equation}
\begin{equation}
\nu_\mu + n \rightarrow \nu_\mu + n +\pi^0
\end{equation}
\begin{equation}
\nu_\mu + n \rightarrow \nu_\mu + p +\pi^- \ .
\end{equation}

The theory for the production of these states is known for thirty years now and there are several calculations available. The charged current reactions have been studied extensively \cite{Adler,Fogli,Rein,Smith,Schreiner,Zucker} and the production of the $\Delta^{++}$ has been understood theoretically. It has also been measured experimentally with good agreement between theory and experiments. For the other charged current reactions there are few experimental measurements. For this reason the nearby detectors of the experiments should study the reactions using light and heavy nuclei as targets.

Knowledge of the neutral current reactions is even more limited. Two of the latest calculations of charged and neutral current reactions for the production of single-pion cross sections differ by approximately $20\%$ \cite{Fogli,Rein}. One should make all possible efforts now to reduce the overall uncertainty and measure the various channels experimentally.

A second difficulty arises from the fact that the experimental targets are heavy materials so that the interactions take place on protons and neutrons bound in nuclei like for example $_8O^{16}$, $_{18}Ar^{40}$ or $_{26}Fe^{56}$. In the heavy nuclei, the produced pions rescatter before they exit from the nucleus and are subject to two phenomena: (1) the cross-sections are reduced by the Pauli exclusion principle, when the energy of the recoiling nucleon is low and can not occupy a filled level of nucleons, and (2) the pion charge exchange due to rescattering. These phenomena are known and have been the subject of extensive studies \cite{Nussinov,Novakowski}.

The expectations of the experiments are the following. For all charge current reactions we expect a reduction of the observed rates in the far away detector because some of the muon neutrinos oscillated into another state. For the neutral current reactions we expect no reduction in rate if the oscillation is to tau neutrino because all neutrinos contribute equally to neutral current reactions. We expect a reduction if the oscillation is to sterile neutrinos.

Now, since the reduction is expected to be approximately $40\%$, it is important to understand all possible corrections. An important requirement is that the nearby and the far away detector use the same nuclei as targets. In case this is not possible, then corrections will have to be applied.

Because of the importance of the experiments and the opportunities they present for establishing the charged and neutral current reactions, we have undertaken the task of calculating the cross-section on free protons and neutrons. This way we produce differential, as well as integrated cross-sections. Then we use the obtained results to calculate the corrections which are present in the nuclei.

This paper is organized as follow: Section 2 is devoted to the calculation of the differential and total cross-sections for single-pion production in neutrino-nucleon interactions. In Section 3 we discuss the nuclear effects involved in this process and calculate the energy spectra for charged and neutral pions for a few different materials typically used as targets in the experiments. Finally, results and conclusions are presented in Section 4.

\section{Single-Pion Production}

\subsection{General Formalism}

In this section we present the main equations and the form factors used to evaluate the differential and total cross sections for single-pion production in neutrino-nucleon interactions. For neutrino energies of a few GeV the single-pion production proceeds mainly through the excitation of the lower resonances. The main contribution to the cross section comes from the production and the subsequent decay of the $\Delta$(1232)P$_{33}$ resonance. Nevertheless, some of the channels receive a non negligible contribution from the isospin 1/2 resonances as, for example, the N(1440)P$_{11}$ and the N(1535)S$_{11}$ resonances.

The channels under investigation in this paper are the three charged current and the four neutral current channels listed in the introduction in Eq. 1.1 to 1.7.

Using Clebsh-Gordan coefficient, it is easy to verify that the amplitudes for these seven channels are given by the following equations:
\begin{eqnarray}
A(\mu^- + p +\pi^+) & = &A_3^{cc}\\ \nonumber 
\\ 
A(\mu^- + n +\pi^+) & = &1/3A_3^{cc} + 2\sqrt2/3A_1^{cc}\\  \nonumber
\\ 
A(\mu^- + p +\pi^0) & = &-\sqrt2/3A_3^{cc} + 2/3A_1^{cc}\\ \nonumber 
\\ 
A(\nu_\mu + p +\pi^0) & = &\sqrt2/3A_3^{nc} + 1/3A_1^{nc} + 1/3A^0_1\\ \nonumber 
\\
A(\nu_\mu + n +\pi^+) & =  &-1/3A_3^{nc} + \sqrt2/3A_1^{nc} + \sqrt2/3A^0_1\\ \nonumber 
\\
A(\nu_\mu + n +\pi^0) & = &\sqrt2/3A_3^{nc} + 1/3A_1^{nc} - 1/3A^0_1\\ \nonumber 
\\
A(\nu_\mu + p +\pi^-) & = &1/3A_3^{nc} - \sqrt2/3A_1^{nc} + \sqrt2/3A^0_1  
\end{eqnarray}
where, in this paper, $A_3^{cc,nc}$ corresponds to the amplitude for the production of the P$_{33}$ resonance, $A_1^{cc,nc}$ is the sum of the amplitudes for the production of the P$_{11}$ and S$_{11}$ resonances and $A^0_1$ is the sum of the isoscalar contributions of the P$_{11}$ and S$_{11}$ resonances to the cross section. 

Notice that, as suggested in Ref.\cite{Fogli}, the neutral current amplitudes $A_3^{nc}$, $A_1^{nc}$ and  $A^0_1$ can be derived from the corresponding charged current amplitudes $A_3^{cc}$ and $A_1^{cc}$ by simply rescaling the vector and axial form factors. In the case of the $A_3^{nc}$ and $A_1^{nc}$ amplitudes the vector and axial charged current form factors need to be multiplied by $1-2sin^2\theta_W$ and by $1$ respectively, where $\theta_W$ is the weak mixing angle. For the  $A^0_1$ instead the vector and axial charged current form factors need to be multiplied respectively by $-2/3sin^2\theta_W$ and $0$. Furthermore, we want to point out that, since the $A^0_1$ amplitude turned out to be very small compared to $A_3^{nc}$ and $A_1^{nc}$, we neglected the isoscalar contribution in our evaluation of the cross sections. 

\subsubsection{$\Delta$(1232)P$_{33}$}

As it was mentioned in the introduction, the theory for the production of the $\Delta$(1232)P$_{33}$ is well known and understood, and several independent calculations have already been published, showing good agreement with the experimental results \cite{Adler,Fogli,Rein,Smith,Schreiner,Zucker}. Therefore, rather than developing our own formalism for this process, we decided to follow the article of Schreiner and von Hippel in Ref.\cite{Schreiner}.

The fully differential cross section ${\rm d}\sigma/{\rm d}Q^2{\rm d}W{\rm d}\Omega_{\pi}$ from Ref.\cite{Schreiner} has been integrated over the polar angle $\phi_{\pi}$ and converted into the triple-differential cross section ${\rm d}\sigma/{\rm d}Q^2{\rm d}W{\rm d}E_{\pi}$ by using the fact that $E^{lab}_{\pi}=\gamma E^{CM}_{\pi}+\beta\gamma|\vec{p}^{CM}_\pi|cos\theta_{\pi}$, where the superscripts $lab$ and $CM$ denote quantities measured in the laboratory and in the center of mass frame respectively. The total cross section has been obtained by integrating over the allowed range of values for $Q^2$ and $E_{\pi}$, with a cut on the invariant mass $W$ range at 1.6 GeV.

The axial and vector form factors used in this calculation are the ones given by Alvarez-Ruso et al. in Eq. 12, 13 and 18  of Ref.\cite{Ruso}. Notice that, since these form factors have been derived from photo- and electro-production experiments in which a $\Delta^+$ or a $\Delta^0$ was produced, in order to obtain the correct cross section for the $\Delta^{++}$ production, all the form factors need to be multiplied by $\sqrt3$ due to the fact that $<\Delta^{++}\mid\rm V_{\alpha}\mid p>=\sqrt3<\Delta^+\mid\rm V^{em}_{\alpha}\mid p>$.

\subsubsection{N(1440)P$_{11}$ and N(1535)S$_{11}$}

As shown in Ref.\cite{Bjorken}, the triple-differential cross section ${\rm d}\sigma/{\rm d}Q^2{\rm d}W{\rm d}E_{\pi}$ for the production of the P$_{11}$ and S$_{11}$ resonances is given by the following equation:
\begin{equation}
{{{\rm d}\sigma}\over{{\rm d}Q^2{\rm d}W{\rm d}E_{\pi}}} = {{{W}\over{4M_N}} {{G_F^2 Q^2}\over{(2\pi)^4\nu^2}}  {{1}\over{\sqrt{\nu^2+Q^2}}}{\left({{\left(1-{{\nu}\over{E_{\nu}}}\right)}|M_s|^2} + {{1\over 2}{{\left(1-{{\nu}\over{E_{\nu}}}\right)}^2|M_r|^2}} + {1\over 2} |M_l|^2\right)}}
\end{equation}
where $M_N$ is the nucleon mass, $G_F$ is the Fermi constant and $\nu$ is the difference between the energies of the incoming and the outgoing lepton. Remembering that nearby the resonance only the s-channel is essential, the three matrix elements $M_i$ with $i = r,l,s$ can be defined as follow:
\begin{equation}
M_i = -f_R\bar{u}(p\prime)\gamma_5(\rlap/p+\rlap/q+M_R)(g_V\rlap/{\epsilon^i}-g_A\rlap/{\epsilon^i}\gamma_5)u(p)f(W)
\end{equation}
where $f_R$ is the coupling constant of the pion to the nucleon and the resonance, $M_R$ is the resonance mass, $p$ and $p\prime$ are the four-momenta of the initial and final nucleon respectively, q is the four-momentum transferred from the leptons to the hadrons and $g_{V,A}$ are the vector and axial form factors. The values for $f_R$ are given in Eq. B7 and B10 in Appendix B of Ref.\cite{Fogli}.

The Breit-Wigner factor $f(W)$ in Eq.2.9 can be written in the following way:
\begin{equation}
{f(W)} = {{1}\over{(W^2-M_R^2)+iM_R\Gamma_R}}
\end{equation}
where $\Gamma_R(W)=\Gamma^0_Rq_{\pi}(W)/q_{\pi}(M_R)$ with $q_{\pi}(W)=\sqrt{(W^2-M_N^2-M_{\pi}^2)^2-4M_N^2M_{\pi}^2}/2W$ and $\Gamma^0_R$ the width of the resonance.

Finally, the polarization vectors $\epsilon^i$ with $i=r,l,s$ used in Eq.2.9 are defined as:
\begin{equation}
{\epsilon^s} = {{1\over Q}{\left(\sqrt{\nu^2+Q^2},0,0,\nu\right)}} \ , \   {\epsilon^{r,l}} = {{1\over\sqrt2}{\left(0,1,\pm i,0\right)}} \ .
\end{equation}
For the form factors we use the expressions obtained by Fogli and Nardulli in Ref.\cite{Fogli}. 

As in the case of the P$_{33}$ resonance, the total cross section has been obtained by integrating over the allowed range of values for $Q^2$ and $E_{\pi}$, with a cut on the invariant mass $W$ range at 1.6 GeV.

\subsection{Results}

In this section we present our results for the total cross section for the seven channels under examination and, where possible, we compare these results with experimental data.

In Fig.1 the total cross section for the $\nu_\mu + p \rightarrow \mu^- + p +\pi^+$ process has been plotted versus the incoming neutrino energy. The data points have been taken from Ref.\cite{Radecky} (solid circles) and from Ref.\cite{Barish} (empty circles). As it can be seen, the agreement between the theoretical curve and the experimental data is quite good.

Fig.2 and Fig.3 display the total cross sections for the $\nu_\mu + n \rightarrow \mu^- + p +\pi^0$ and $\nu_\mu + n \rightarrow \mu^- + n +\pi^+$ processes respectively, again plotted versus the incoming neutrino energy. In this case the data points have been taken from Ref.\cite{Radecky} (solid circles), from Ref.\cite{Barish} (empty circles) and from Ref.\cite{Grabosch} (crosses). Also in this case the agreement between the theoretical results and the data points is reasonably good.

The difference between the theoretical and the experimental results can be partially explained by taking into account the fact that, while the theoretical curves have been estimated imposing a cut on the invariant mass $W$ at 1.6 GeV, the experimental points have been obtained without any cut. Notice also that we didn't include any non-resonant background in our evaluation of the cross sections.

In the case of the neutral current interactions, the experimental results are presented in the form of ratios between each of the neutral current channels and one of the charged current channels. For these reason, Fig.4, Fig.5, Fig.6 and Fig.7, which display respectively the total cross sections of the $\nu_\mu + p \rightarrow \nu_\mu + p +\pi^0$, $\nu_\mu + p \rightarrow \nu_\mu + n +\pi^+$, $\nu_\mu + n \rightarrow \nu_\mu + n +\pi^0$ and $\nu_\mu + n \rightarrow \nu_\mu + p +\pi^-$ processes plotted versus the incoming energy, have no data points. Nevertheless, we compared our results with the experimental ratios from Ref.\cite{Krenz,Derrick,Barish-r,Lee} and found that there is a reasonable agreement, even if, in some cases, the ratios measured by the different experiments differ a lot one from the other.

\section{Nuclear Effects}

In Section 2 we discussed the reaction $\nu + N \rightarrow l + N' +\pi^{\pm,0}$, where $N$ is a free nucleon (proton or neutron). In order to investigate the nuclear effects taking place in the experimental targets (for example, $_8O^{16}$, $_{18}Ar^{40}$ or $_{26}Fe^{56}$), we need to study the modifications necessary for the reaction $\nu + T \rightarrow l + T' +\pi^{\pm,0}$, where $T$ is the nuclear target and $T'$ is an unobserved final nuclear state.

We visualize the reaction as a two step process with the neutrino interacting with individual nucleons producing single pions and excited nuclei. The production process is influenced by the Pauli principle and the Fermi-motion of individual nucleons. The subsequent journey of the pions is a ``random-walk'' of multiple scattering until the pion escape from the nucleus. In the multiple scattering the pions can exchange their charge. These phenomena have been studied in the past \cite{Nussinov} and we shall adopt the formalism in order to calculate the energy spectra of the pions. We give enough details so that the reader has an overview of the model but for more details he or she should consult the original article.

\subsection{Charge Density Distributions}

Following Ref.\cite{Nussinov}, we treat the target as a collection of independent nucleons which are distributed in space accordingly to a density profile determined through electron-nucleus scattering experiments. For the charge density profile of $_8O^{16}$ we adopt the harmonic oscillator model in which the density is given by:
\begin{equation}
{\rho(r)} = {\rho(0)\exp(-r^2/R^2)\left(1+C{r^2\over R^2}+C_1{\left(r^2\over R^2\right)^2}\right)}
\end{equation}
where $R = a/K$ with $K = \sqrt{3(2+5C)/2(2+3C)}$ and $a$ the root mean square radius.

For $_{18}Ar^{40}$ and $_{26}Fe^{56}$ we use the two parameters Fermi model and write the charge density in the following way:
\begin{equation}
{\rho(r)} = {\rho(0)\left(1+\exp{((r-C)/C_1)}\right)^{-1}} \ .
\end{equation}

The different parameters used in Eq.[3.1,3.2] are given in Ref.\cite{Vries} and are summarized in Table 1.

\subsection{Charge Exchange}

In the first step the neutrinos interact with the bound protons and neutrons with these reactions allowed provided that the energy of the recoiling nucleon is above the Fermi sea. This brings a correction factor calculated in \cite{Nussinov}. In the second step the pions rescatter several times until they reach the surface of the nucleus and escape. It is important to notice that the process taking place during the rescattering depend only on the properties of the target nucleus and are independent of the leptons involved in the first step.
The differential cross sections for leptonic pion production on nuclear and on free nucleon targets are related to each other through the so called charge exchange matrix $M$ in the following way:
\begin{equation}
{\left(\begin{array}{c}\displaystyle
{{\rm d}\sigma(_ZT^A;+)\over {\rm d}Q^2{\rm d}W{\rm d}E_{\pi}}\\
\displaystyle{{\rm d}\sigma(_ZT^A;0)\over {\rm d}Q^2{\rm d}W{\rm d}E_{\pi}}\\
\displaystyle{{\rm d}\sigma(_ZT^A;-)\over {\rm d}Q^2{\rm d}W{\rm d}E_{\pi}}
\end{array}\right)} ={ M {\left(\begin{array}{c}\displaystyle
{{\rm d}\sigma(N_T;+)\over {\rm d}Q^2{\rm d}W{\rm d}E_{\pi}}\\
\displaystyle{{\rm d}\sigma(N_T;0)\over {\rm d}Q^2{\rm d}W{\rm d}E_{\pi}}\\
\displaystyle{{\rm d}\sigma(N_T;-)\over {\rm d}Q^2{\rm d}W{\rm d}E_{\pi}}
\end{array}\right)}}
\end{equation}
where 
\begin{equation}
{{\rm d}\sigma(N_T;\pm 0)\over {\rm d}Q^2{\rm d}W{\rm d}E_{\pi}} = {Z{{\rm d}\sigma(p;\pm 0)\over {\rm d}Q^2{\rm d}W{\rm d}E_{\pi}}+(A-Z){{\rm d}\sigma(n;\pm 0)\over {\rm d}Q^2{\rm d}W{\rm d}E_{\pi}}} \ .
\end{equation}

 Its eigenvalues define beams of pions of specific charge combination, which offer a scattering they produce a beam, which is decreased by the appropriate eigenvalue. The complete scattering phenomenon is characterized by three transition probabilities f($\lambda$) corresponding to the three eigenvalues. They describe the probabilities of beams with eigenvalues $\lambda$ = 1, $\frac{5}{6}$ and $\frac{1}{2}$ to survive and exit the nucleus. The ``random-walk'' of the pions is a stochastic process and several solutions were found in Ref.\cite{Nussinov}. We collect here the main  formulas for the calculation. When we transpose the final pion to the initial state, we obtain the system $\pi_i+\bar{\pi}_f$ whose total isospin can be 0,1 and 2. We use this property to parameterize the charge exchange matrix in terms of three functions A, c and d   
\begin{equation}
M = A \left(\begin{array}{ccc}
{1-c-d} & d & c \\
d & {1-2d} & d \\
c & d & {1-c-d}
\end{array}\right) \ .
\end{equation}
The overall factor A is given by
\begin{equation}
A = g(W,Q^2)f(1)
\end{equation}
with g(W,$Q^2$) being the Pauli suppression factor and f(1) the transmission coefficient for the state with eigenvalue 1. The others two functions are
\begin{equation}
c = {1\over 3}-{1\over 2}{f(5/6)\over f(1)}+{1\over 6}{f(1/2)\over f(1)}
\end{equation}
\begin{equation}
d = {1\over 3}{\left(1-{f(1/2)\over f(1)}\right)} \ .
\end{equation}

As mentioned already f($\lambda$) contains the dynamics of the multiple scattering for the $\lambda$ eigenvalues. Both the Pauli factor and several models f($\lambda$) were presented in Ref.\cite{Nussinov}.
It was also shown that approximating the multiple-scattering with scatterings in the forward and backward directions provides a very accurate approximation. Thus what is important is the effective profile of the nucleus that the pions sea. This allows one to write   
\begin{equation}
{f(\lambda)} = {{\int_0^{\infty} {\rm d}b \,bL(b)f(\lambda,L(b))}\over {\int_0^{\infty} {\rm d}b \,bL(b)}}
\end{equation}
where $b$ is the impact parameter and the effective length $L(b)$ is given by:
\begin{equation}
{L(b)} = {{1\over \rho(0)}{\int_{-\infty}^{+\infty} {\rm d}z \,\rho\sqrt{(z^2+b^2)}}} \ .
\end{equation}
In the case of $_8O^{16}$, the effective length $L(b)$ is given by:
\begin{equation}
{L(b)} = {R\sqrt{\pi}\exp{-b^2/R^2}{\left(1+C\left({1\over 2}+{b^2\over R^2}\right)\right)}}
\end{equation} 
while for $_{18}Ar^{40}$ and $_{26}Fe^{56}$ $L(b)$ is written as: 
\begin{equation}
{L(b)} = {{2\over 3}{\left(C\over C_1\right)^3}{\left(\pi^2{\left(C\over C_1\right)}+{\left(C\over C_1\right)^3}-6\sum_{n=1}^{\infty}{{\left(-\exp(-x)\right)^n}\over {n^3}}\right)}} \ .
\end{equation} 
The appropriate expression for the function $f(\lambda,L(b))$ has been derived in Ref.\cite{Nussinov} both for the case of a one-dimensional multiple scattering problem and for the case of a spherical one. The two solutions have then been compared showing excellent agreement over the entire range of parameters. Therefore, in this paper, we adopted for $f(\lambda,L(b))$ the approximate expression obtained in Ref.\cite{Nussinov} for the one-dimensional problem.

\subsection{Averaging Approximation}

It is important to notice that, while the Pauli production factor depends on both $W$ and $Q^2$, the function $f(\lambda)$ depends only on W and this dependence is very weak. Therefore, as it has been already verified in Ref.\cite{Nussinov}, it is reasonable to average the charge exchange parameters over the leading $W$-dependence by defining an averaged function $\bar{f}(\lambda)$ in the following way:
\begin{equation}
{\bar{f}(\lambda)} = {{\int {\rm d}W \,q(W)^{-1}\sigma_{3,3}(W)f(\lambda)}\over {\int {\rm d}W \,q(W)^{-1}\sigma_{3,3}(W)}}
\end{equation}
where $\sigma_{3,3}(W)$ is the pion-nucleon scattering cross section and $q(W)$ is the pion momentum. For the definitions of $\sigma_{3,3}(W)$ and $q(W)$ see Appendix C of Ref.\cite{Nussinov}.

\subsection{Results}

Using the model outlined in the previous subsections, we evaluated the nuclear corrections for leptonic pion production on three different nuclei: oxygen, argon and iron.

The values obtained for $f(\lambda)$ and for $g(W,Q^2)$ for the three different targets under consideration are listed in Table II and Table III, respectively. Notice that, as Table III shows, the reduction of the cross section due to Pauli exclusion principle is larger for smaller values of $Q^2$ and it does not depend on the material.

The charge exchange matrices M for oxygen, argon and iron are given by:
\begin{equation}
M(_8O^{16}) = A \left(\begin{array}{ccc}
0.782 & 0.161 & 0.057 \\
0.161 & 0.677 & 0.161 \\
0.057 & 0.161 & 0.782
\end{array}\right)
\end{equation}
\begin{equation}
M(_{18}Ar^{40}) = A \left(\begin{array}{ccc}
0.731 & 0.187 & 0.082 \\
0.187 & 0.625 & 0.187 \\
0.082 & 0.187 & 0.731
\end{array}\right)
\end{equation}
\begin{equation}
M(_{26}Fe^{56}) = A \left(\begin{array}{ccc}
0.718 & 0.194 & 0.088 \\
0.194 & 0.612 & 0.194 \\
0.088 & 0.194 & 0.718
\end{array}\right) \ .
\end{equation}

The Pauli factor and the charge exchange matrix M for oxygen have been compared with the corresponding quantities previously evaluated in Ref.\cite{Nussinov,Novakowski} and have found to be in good agreement with each other. Unfortunately, no comparison with previous calculation or experimental data is possible for argon and iron.

The differential cross sections evaluated in Sect. 2 for free nucleon targets have been used here together with the charge exchange matrix $M$ to obtain the differential cross sections for nuclear targets. These cross sections have been integrated over $W$ and $Q^2$ keeping the neutrino energy fix at 1 GeV in order to obtain the pion energy spectra appearing in Fig.[8-16].

Fig.[8-10] display respectively the pion energy distributions for positive, neutral and negative pions produced on oxygen targets. In each figure the solid line represent the initial distribution without any nuclear correction, the dashed line represents the same distribution after the application of the Pauli factor in the production, and the dotted line represents the final distribution after applying all the nuclear corrections discussed in the previous subsections. Similarly, Fig.[11-13] and Fig.[14-16] display the corresponding pion distributions produced on argon and iron targets, respectively.

From these figures it is clear that, while the reduction of the cross section due to the Pauli production factor is the same for all the processes investigated in this paper, the nuclear corrections related to the pion charge exchange and pion absorption are larger for neutral pions than for the positive or negative ones. Furthermore, these corrections turn out to be larger for heavier nuclei. Finally, the magnitude of the nuclear corrections decreases with increasing pion energy.

\section{Conclusions}
At least three long-baseline neutrino experiments plan to study low energy neutrino reactions. Their main aim is the observation and better understanding of the neutrino-oscillations, but a necessary input is the understanding of these reactions in free protons and neutrons, as well as the modifications brought about when the nucleons are bound in relatively heavy nuclei.

In order to work in a coherent framework we calculated the cross sections on free protons and neutrons. The theory for the production of the $\Delta$(1232)P$_{33}$ resonance is well understood and our results for the total cross section agree with the experimental data. The same holds for other channels where I=1/2 resonances also contribute. The comparisons appear in Figures 1-3, where it is evident that the accuracy of the measurements is subject to large improvements. Thus it is highly desirable that the new experiments use the nearby detector in order to measure the various cross sections. This refers to charged and neutral currents interactions on free protons and neutrons. The main uncertainties on this part of the paper are the functional form and parameters of the form factors and interference between I=3/2 and I=1/2 resonances. We expect that the effects from these uncertainties are small.

More important are changes which are brought about in the scattering of neutrinos in heavy nuclei. It is very likely that the far away detectors will use heavy materials as targets in order to enhance their counting rates. The  heavy materials bring in corrections comparable to oscillations. In this article we used an old model for nuclear corrections \cite{Nussinov} and calculated the effects on the produced $\pi^{\pm ,0}$. In Section 3 we reviewed the main features of the model so that the interested reader can reproduce the results. 

We decided that an interesting and important parameter in the experiments is the energy of the emerging pion. We calculated in Figures 8-16 the pion spectra as function of their energy. We found that the largest correction appears in the spectrum of the $\pi ^0$'s. The reduction of the signal for neutral pions with energies approximately 200 MeV is substantial: of the order of 40$\%$. Processes with nuclear corrections as large as the ones found in this article require special attention. Several strategies suggest themselves.

One is to use the same material for the front and the far away detector and study the spectra as a function of $E_{\pi}$. Then compare the results from the two detectors and with quasi elastic scattering. In case that the experiments are forced to use different materials detailed calculations for the two materials will point to similarities and possible differences between the two targets. 

\acknowledgements
In the progress of this work we profited from the expertise of our colleagues. We wish to thank S. L. Adler, D. Rein, M. H. Reno, J. F. J. Salgado and L. Sehgal for helpful discussions. One of us (EAP) thanks S. L. Adler and the Institute for Advanced Study for its hospitality where part part of this work was performed. The work of LP is supported by the Deutsche Forschungsgemeinschaft (DFG) under contract GRK 54/3. The work of JYY is supported by the German Federal Ministry of Science (BMBF) under contract 05HT9PEA5.

% TABLES

%table 1
\begin{table}
\caption{Charge density distribution paprameters}
\begin{tabular}{cccccc}
$_ZT^A$ & $a[fm]$ & $C[fm]$ & $C_1[fm]$ & $R[fm]$ & $\rho(0)[fm^{-3}]$ \\ \hline
$_{8}O^{16}$ & 2.718 & 1.544 & 0 & 1.833 & 0.141\\
$_{18}Ar^{40}$ & 3.393 & 3.530 & 0.542 & 4.830 & 0.176\\
$_{26}Fe^{56}$ & 3.801 & 4.111 & 0.558 & 4.907 & 0.163\\
\end{tabular}
\end{table}

%table 2
\begin{table}
\caption{$\bar{f}(\lambda)$ with $\lambda$ = 1, 1/2 and 5/6 for $_{8}O^{16}$, $_{18}Ar^{40}$ and $_{26}Fe^{56}$.}
\begin{tabular}{cccc}
$_ZT^A$ & $\bar{f}(1)$ & $\bar{f}(1/2)$ & $\bar{f}(5/6)$  \\ \hline
$_{8}O^{16}$ & 0.811 & 0.418 & 0.587 \\
$_{18}Ar^{40}$ & 0.648 & 0.284 & 0.420 \\
$_{26}Fe^{56}$ & 0.625 & 0.261 & 0.393 
\end{tabular}
\end{table}

%table 3
\begin{table}
\caption{Pauli production factor $g(W,Q^2)$ evaluated for $W=1.2 GeV$.}
\begin{tabular}{cc}
$Q^2$ & $g(1.2 GeV,Q^2)$ \\ \hline
0.00 & 0.785 \\
0.05 & 0.872 \\
0.10 & 0.926 \\
0.15 & 0.960 \\
0.20 & 0.980 \\
0.25 & 0.992 \\
0.30 & 0.997 \\
0.35 & 1.000 \\
0.40 & 1.000 \\
\end{tabular}
\end{table}

% FIGURES

%figure 1
\begin{figure}
\centerline{\psfig{figure=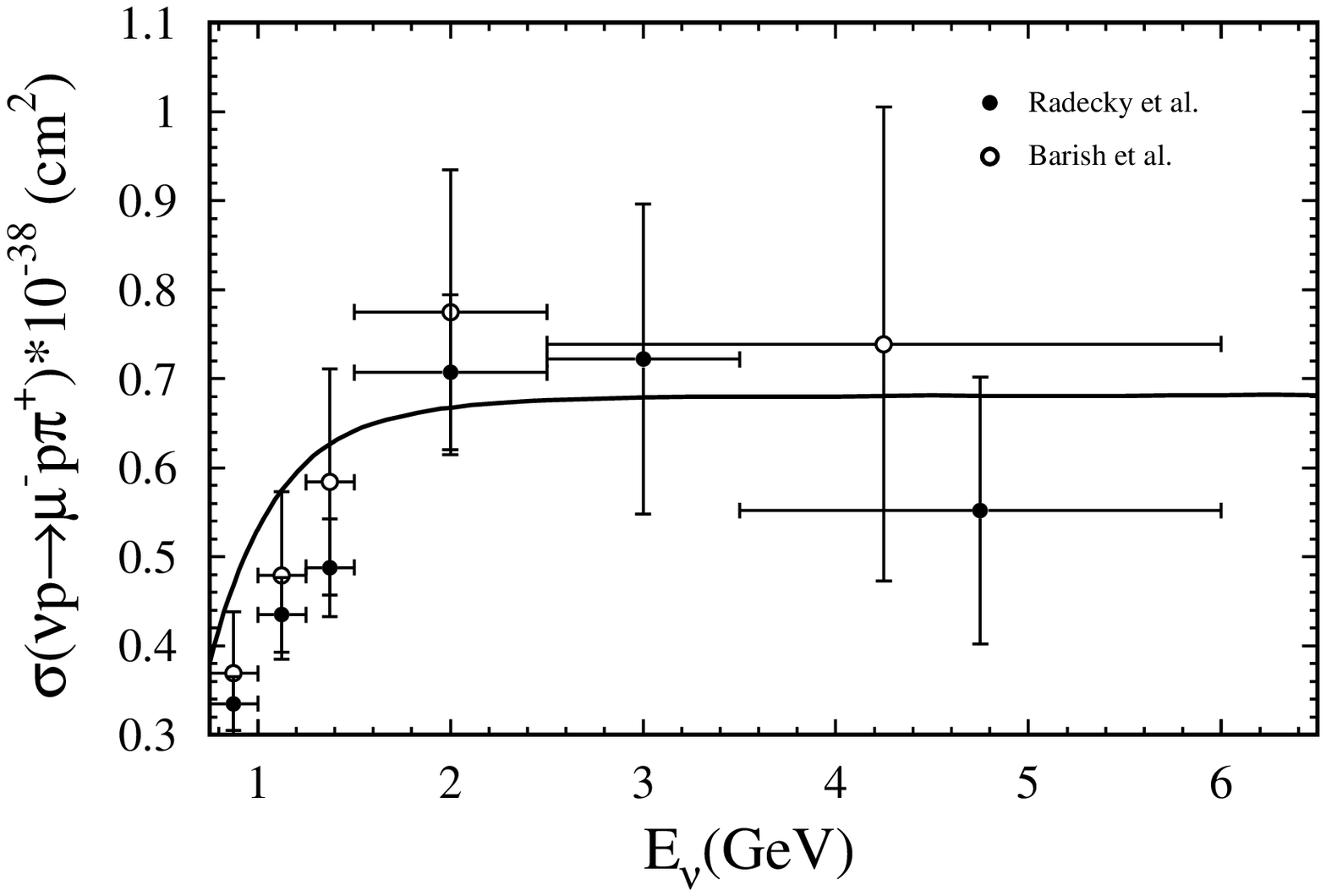,height=3.5in,angle=0}}
\caption{Total cross section for the $\nu_\mu + p \rightarrow \mu^- + p +\pi^+$ process plotted versus the incoming neutrino energy. The data points are from Ref.[13] and Ref.[14].}
\end{figure}

%figure 2
\begin{figure}
\centerline{\psfig{figure=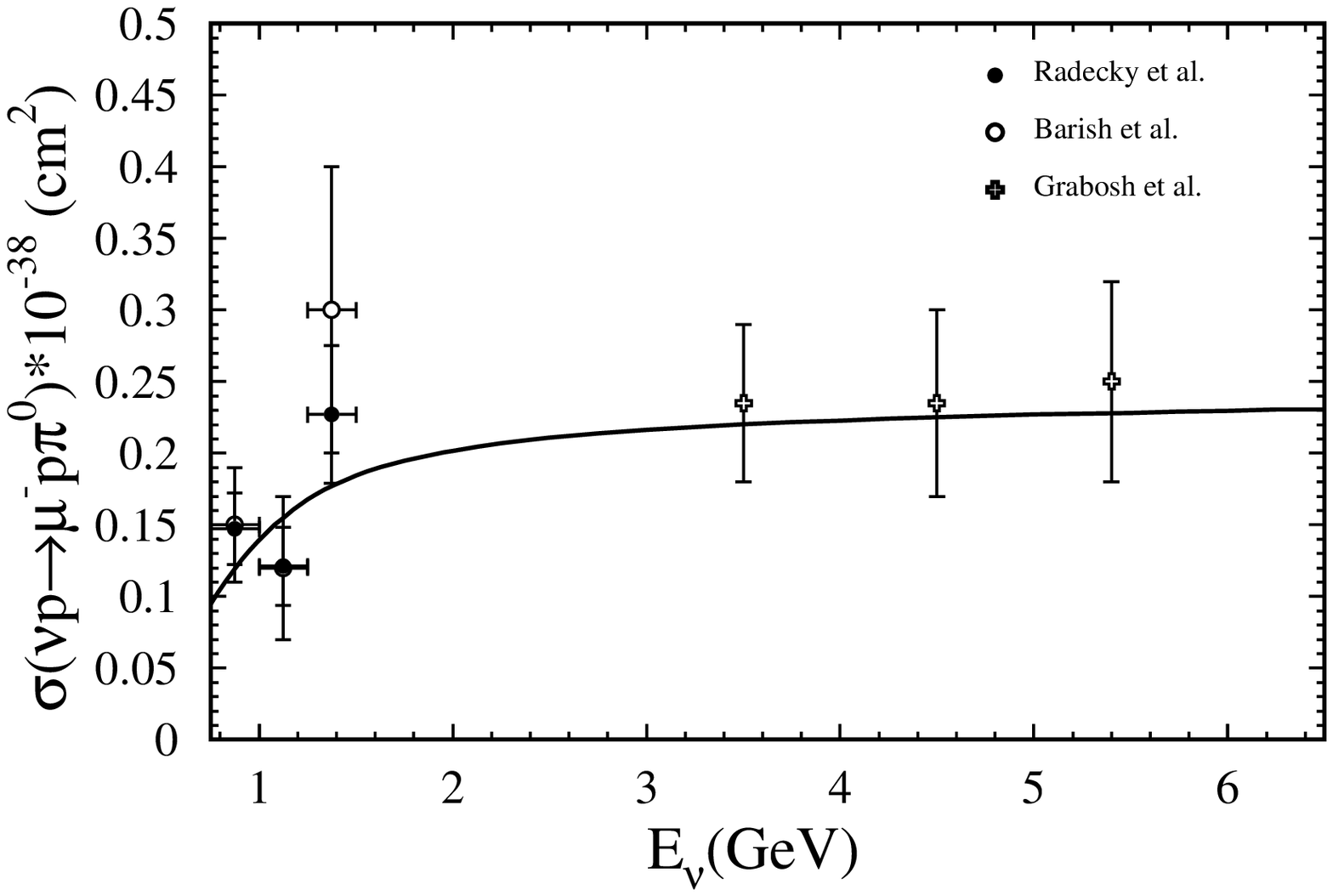,height=3.5in,angle=0}}
\caption{Total cross section for the $\nu_\mu + n \rightarrow \mu^- + p +\pi^0$ process plotted versus the incoming neutrino energy. The data points are from Ref.[13], Ref.[14] and Ref.[15].}
\end{figure}

%figure 3
\begin{figure}
\centerline{\psfig{figure=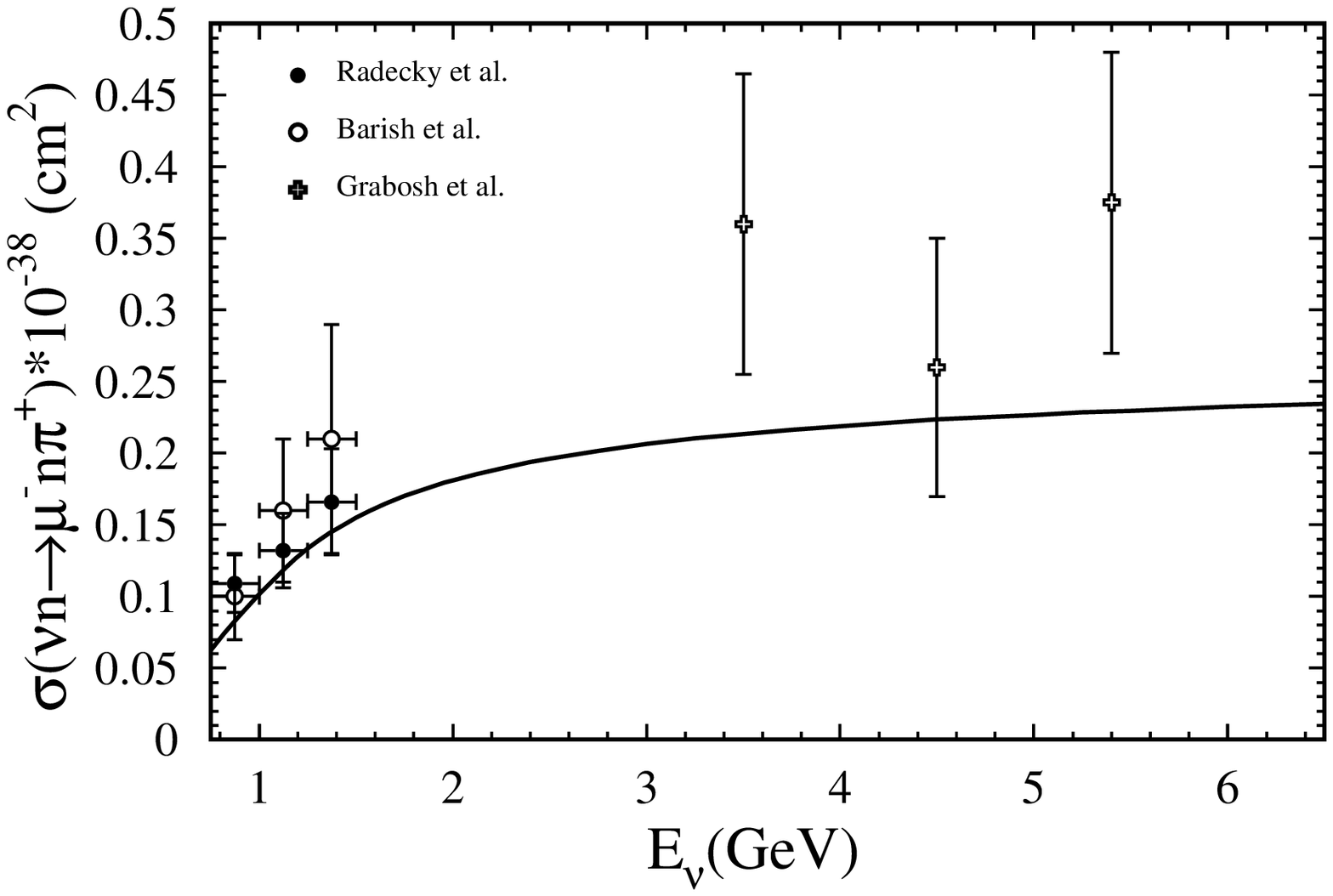,height=3.5in,angle=0}}
\caption{Total cross section for the $\nu_\mu + n \rightarrow \mu^- + n +\pi^+$ process plotted versus the incoming neutrino energy. The data points are from Ref.[13], Ref.[14] and Ref.[15].}
\end{figure}

%figure 4
\begin{figure}
\centerline{\psfig{figure=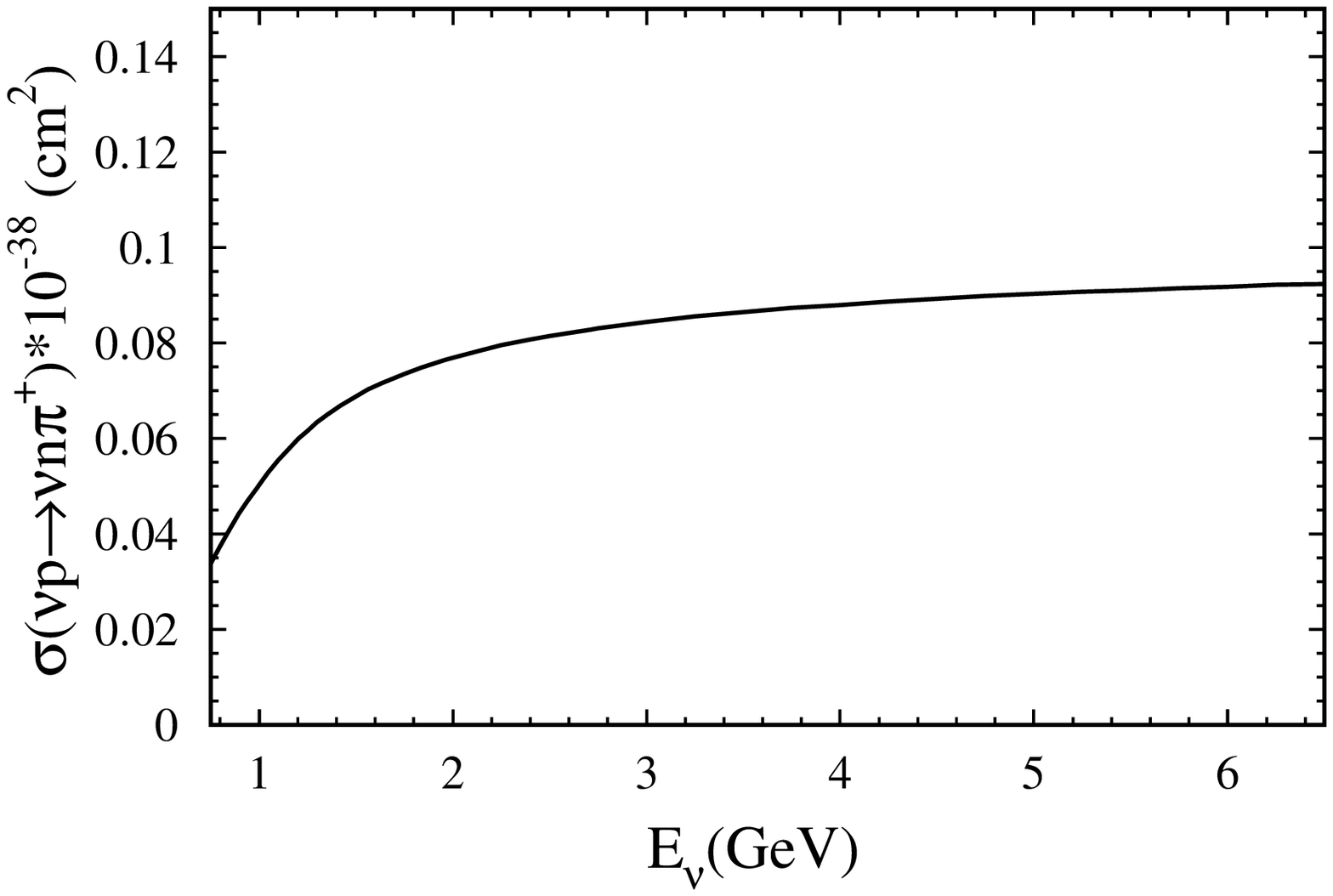,height=3.5in,angle=0}}
\caption{Total cross section for the $\nu_\mu + p \rightarrow \nu_\mu + n +\pi^+$ process plotted versus the incoming neutrino energy.}
\end{figure}

%figure 5
\begin{figure}
\centerline{\psfig{figure=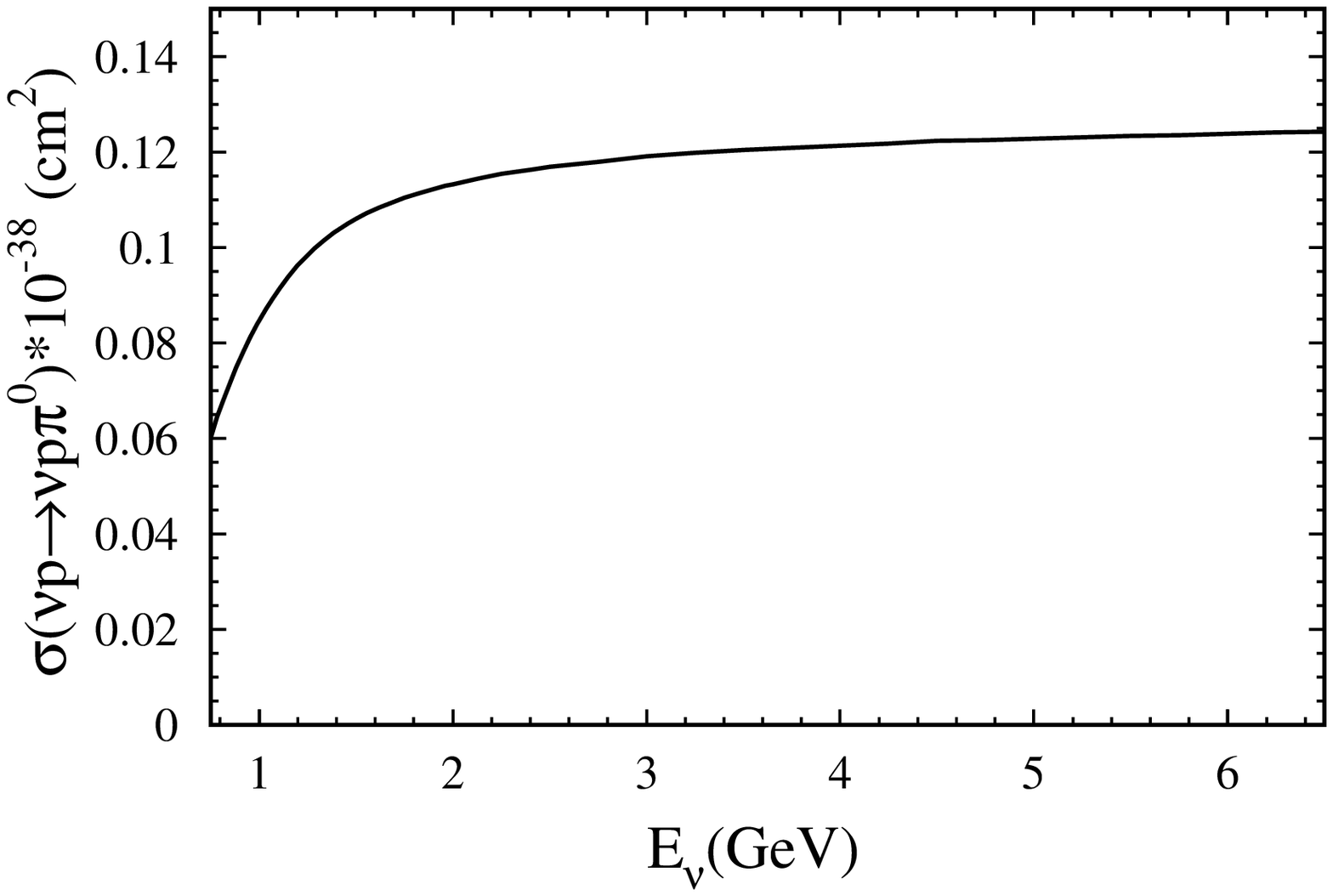,height=3.5in,angle=0}}
\caption{Total cross section for the $\nu_\mu + p \rightarrow \nu_\mu + p +\pi^0$ process plotted versus the incoming neutrino energy. }
\end{figure}

%figure 6
\begin{figure}
\centerline{\psfig{figure=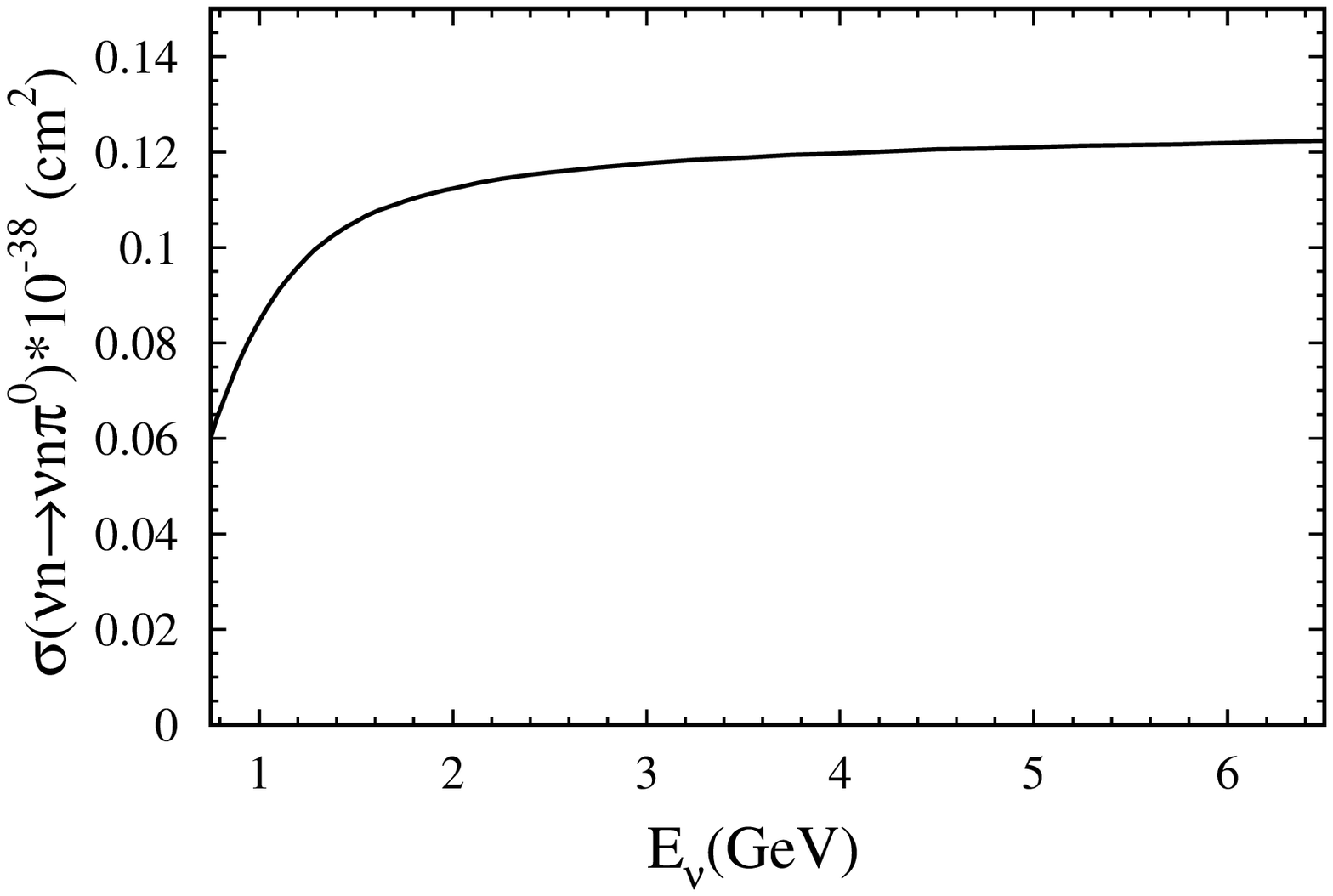,height=3.5in,angle=0}}
\caption{Total cross section for the $\nu_\mu + n \rightarrow \nu_\mu + n +\pi^0$ process plotted versus the incoming neutrino energy.}
\end{figure}

%figure 7
\begin{figure}
\centerline{\psfig{figure=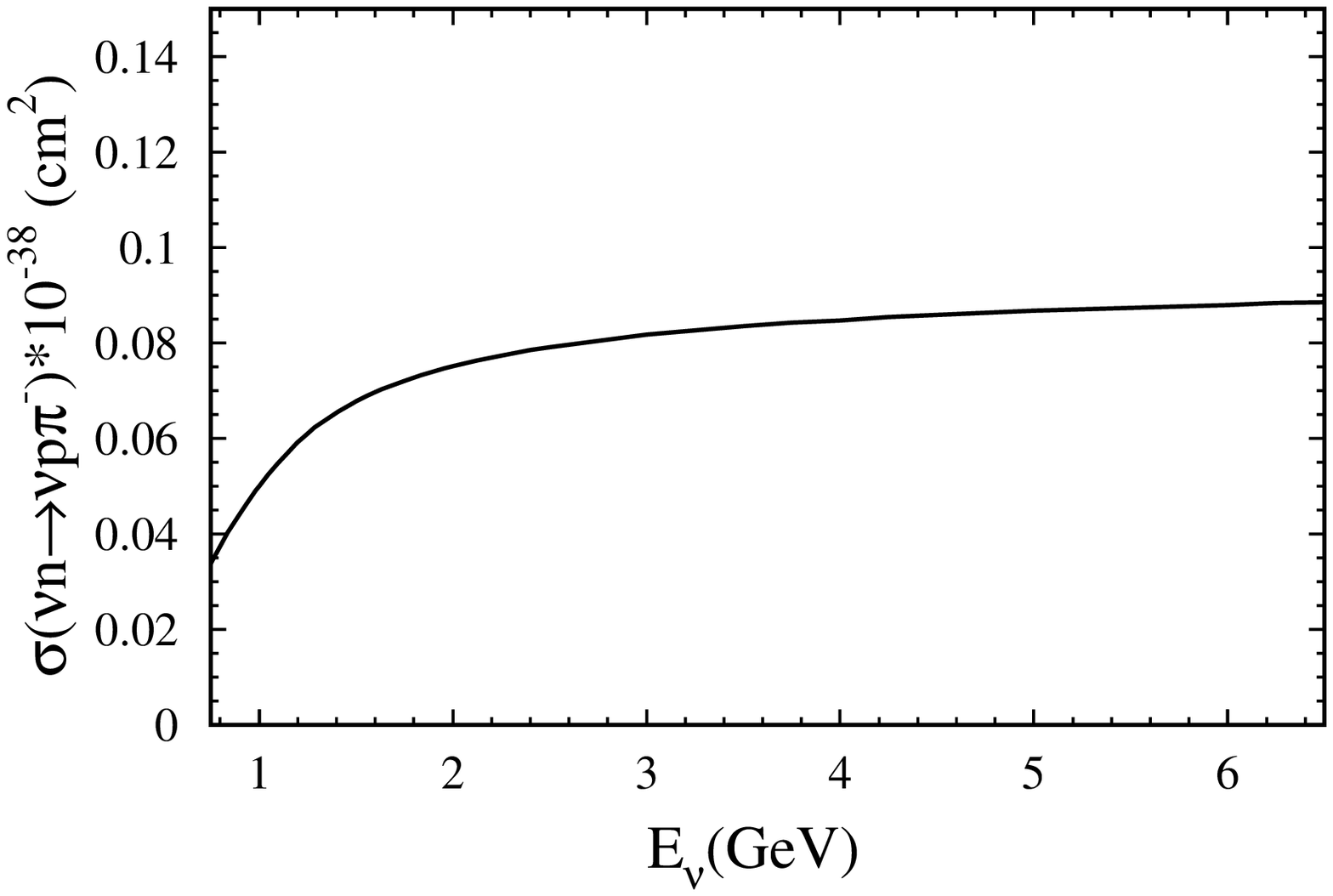,height=3.5in,angle=0}}
\caption{Total cross section for the $\nu_\mu + n \rightarrow \nu_\mu + p +\pi^-$ process plotted versus the incoming neutrino energy.}
\end{figure}

%figure 8
\begin{figure}
\centerline{\psfig{figure=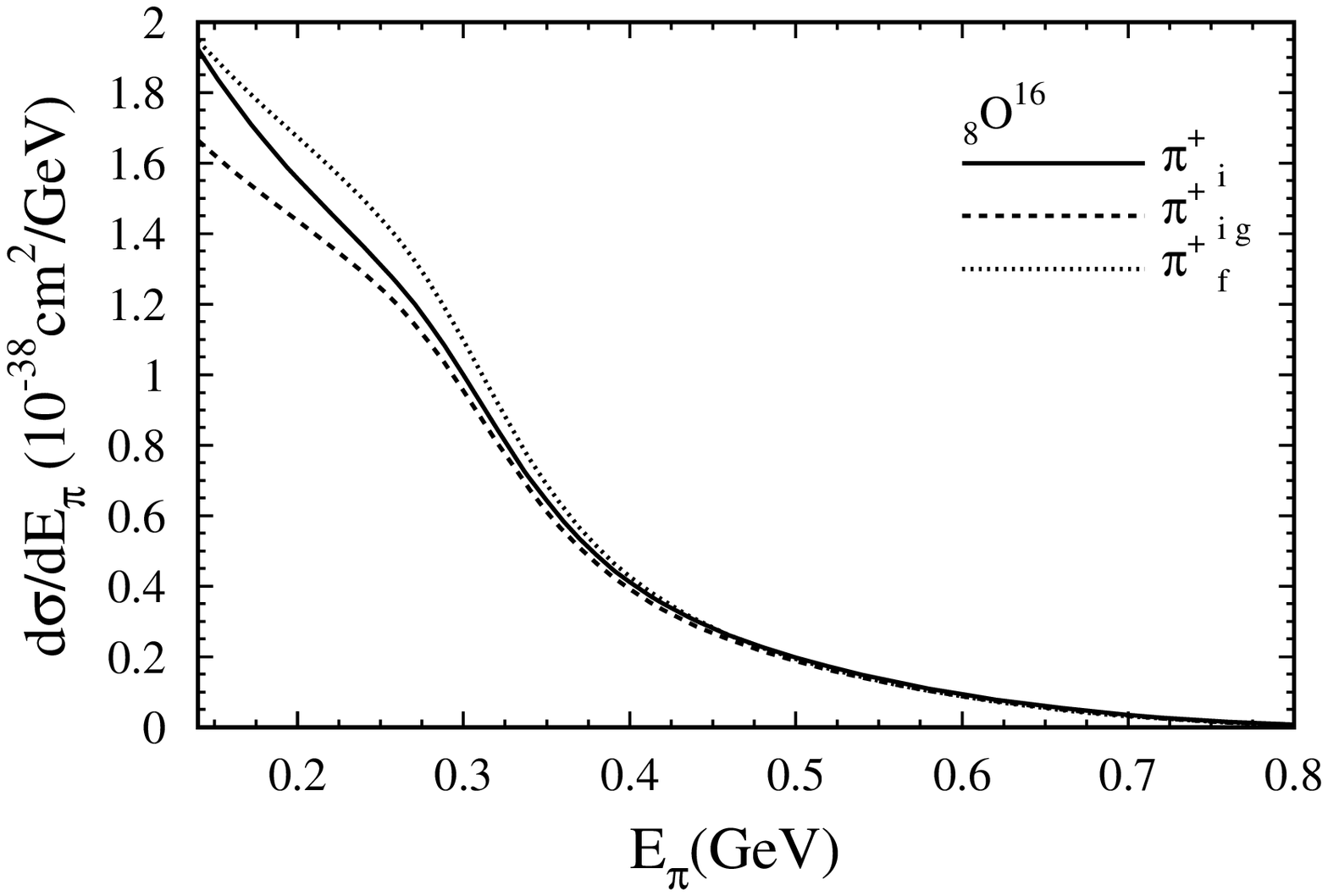,height=3.5in,angle=0}}
\caption{Pion energy distribution for positively charged pions produced on oxygen targets. The solid, dashed and dotted lines represent respectively the pion energy distribution without any nuclear correction, including only the Pauli production factor $g$ and including all nuclear corrections.}
\end{figure}

%figure 9
\begin{figure}
\centerline{\psfig{figure=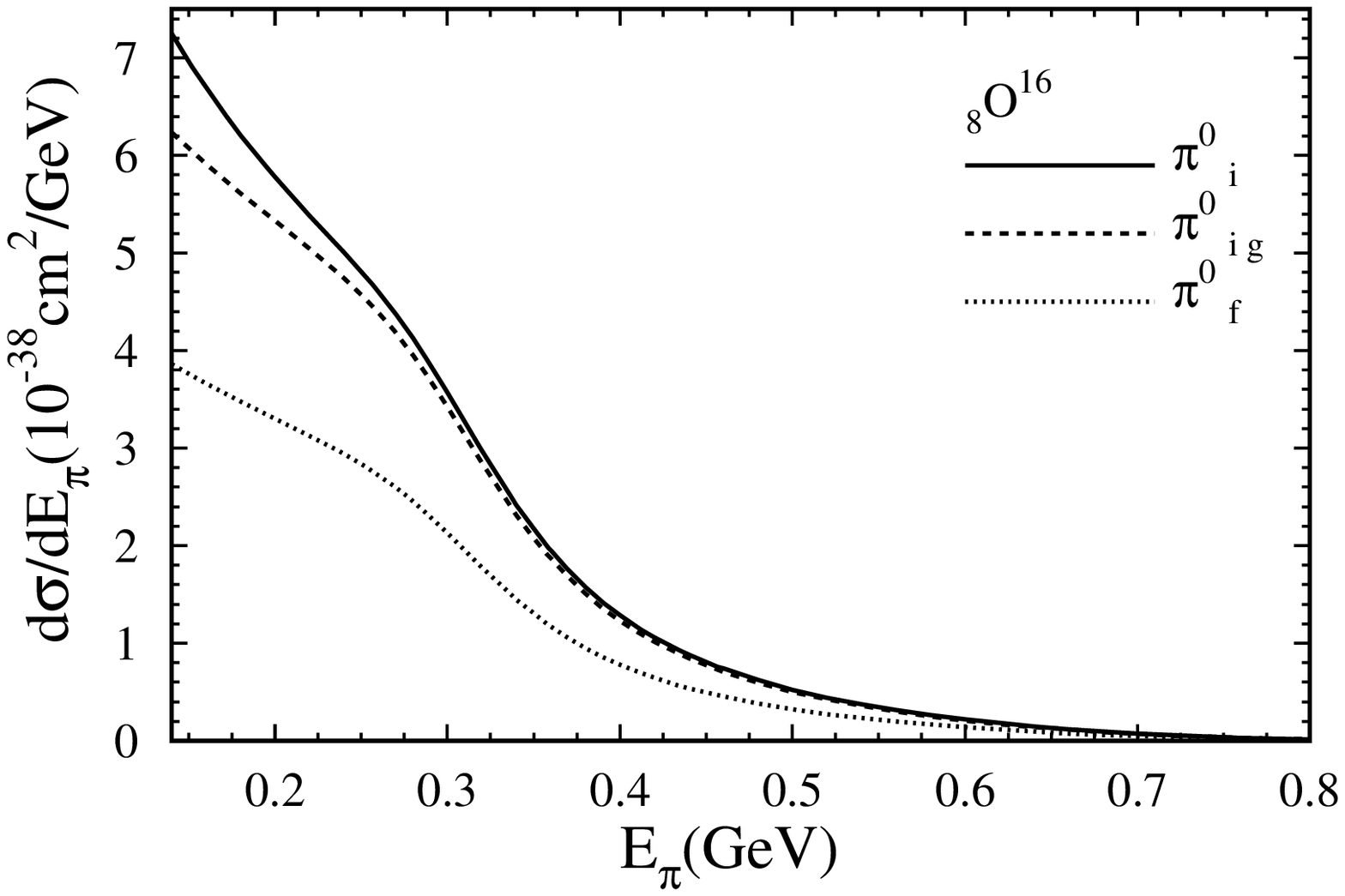,height=3.5in,angle=0}}
\caption{Pion energy distribution for neutrally charged pions produced on oxygen targets. The solid, dashed and dotted lines represent respectively the pion energy distribution without any nuclear correction, including only the Pauli production factor $g$ and including all nuclear corrections.}
\end{figure}

%figure 10
\begin{figure}
\centerline{\psfig{figure=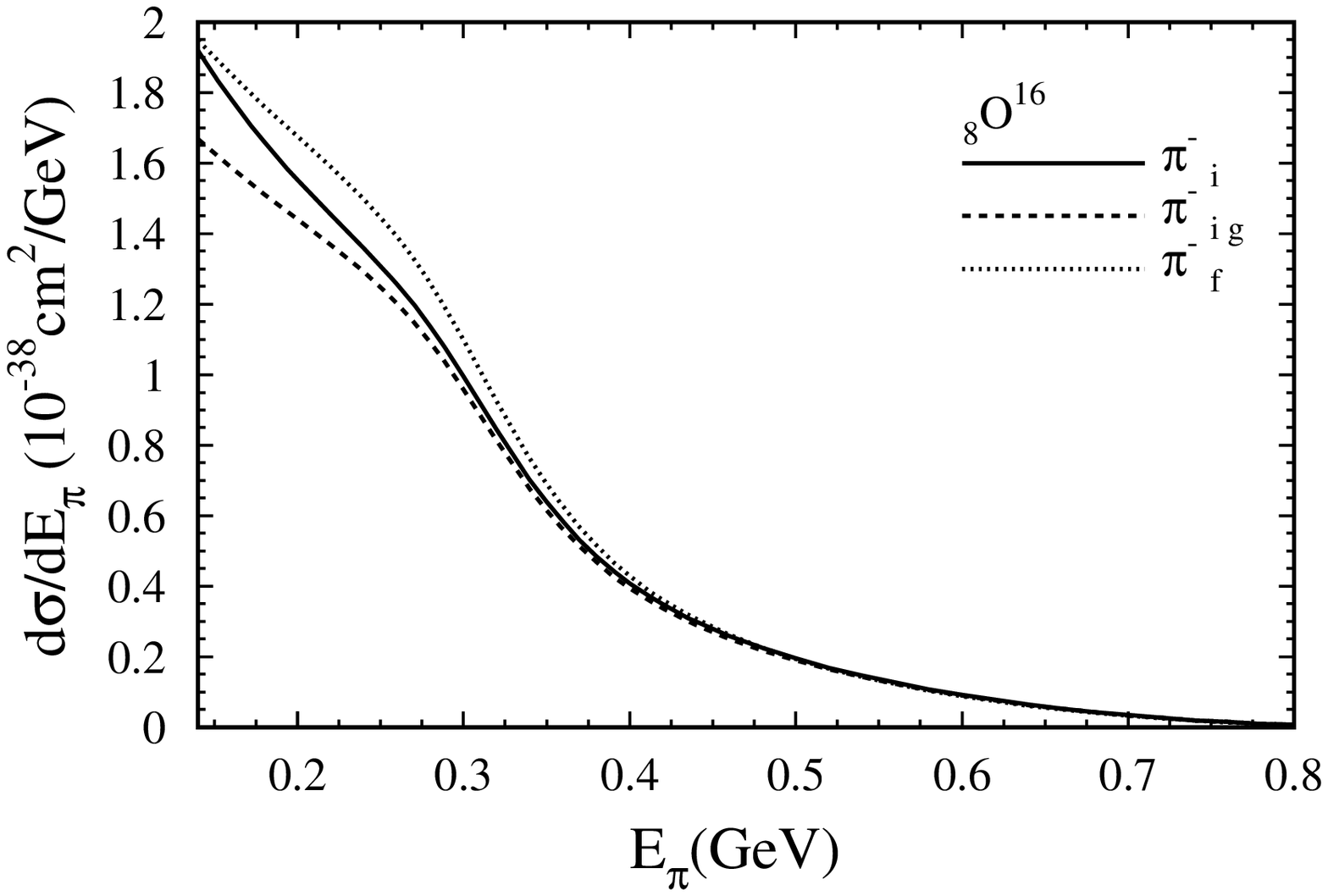,height=3.5in,angle=0}}
\caption{Pion energy distribution for negatively charged pions produced on oxygen targets. The solid, dashed and dotted lines represent respectively the pion energy distribution without any nuclear correction, including only the Pauli production factor $g$ and including all nuclear corrections.}
\end{figure}

%figure 11
\begin{figure}
\centerline{\psfig{figure=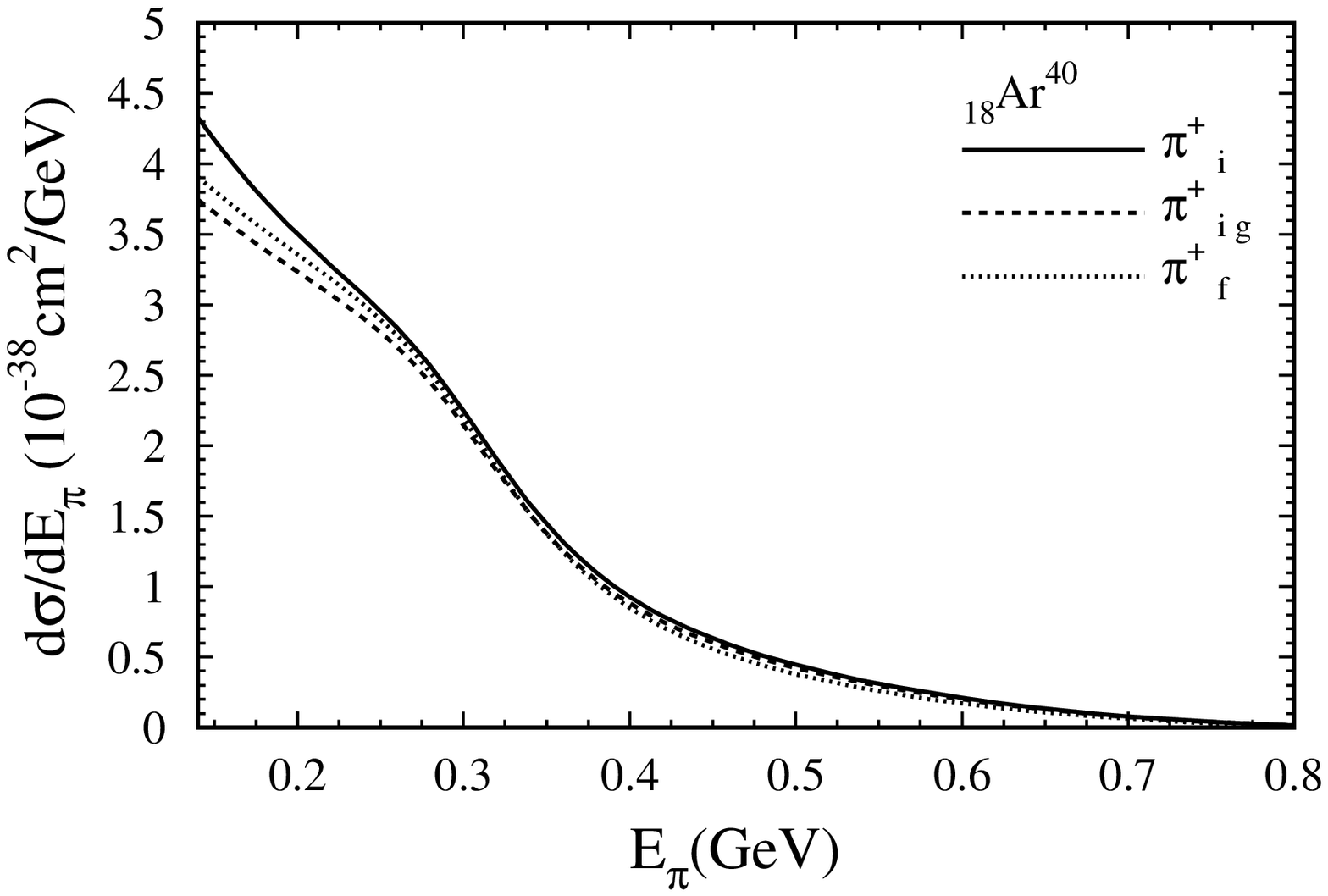,height=3.5in,angle=0}}
\caption{Pion energy distribution for positively charged pions produced on argon targets. The solid, dashed and dotted lines represent respectively the pion energy distribution without any nuclear correction, including only the Pauli production factor $g$ and including all nuclear corrections.}
\end{figure}

%figure 12
\begin{figure}
\centerline{\psfig{figure=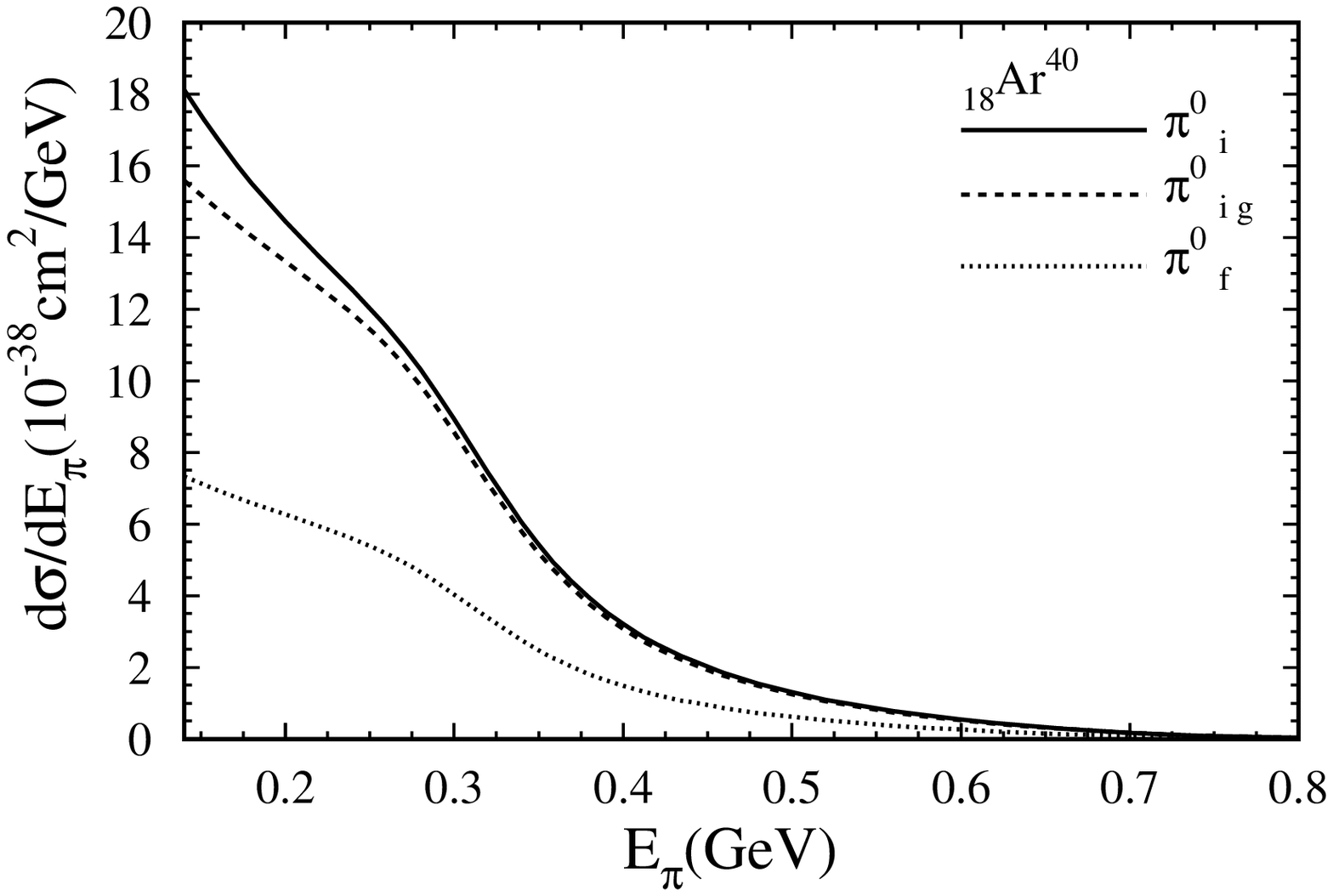,height=3.5in,angle=0}}
\caption{Pion energy distribution for neutrally charged pions produced on argon targets. The solid, dashed and dotted lines represent respectively the pion energy distribution without any nuclear correction, including only the Pauli production factor $g$ and including all nuclear corrections.}
\end{figure}

%figure 13
\begin{figure}
\centerline{\psfig{figure=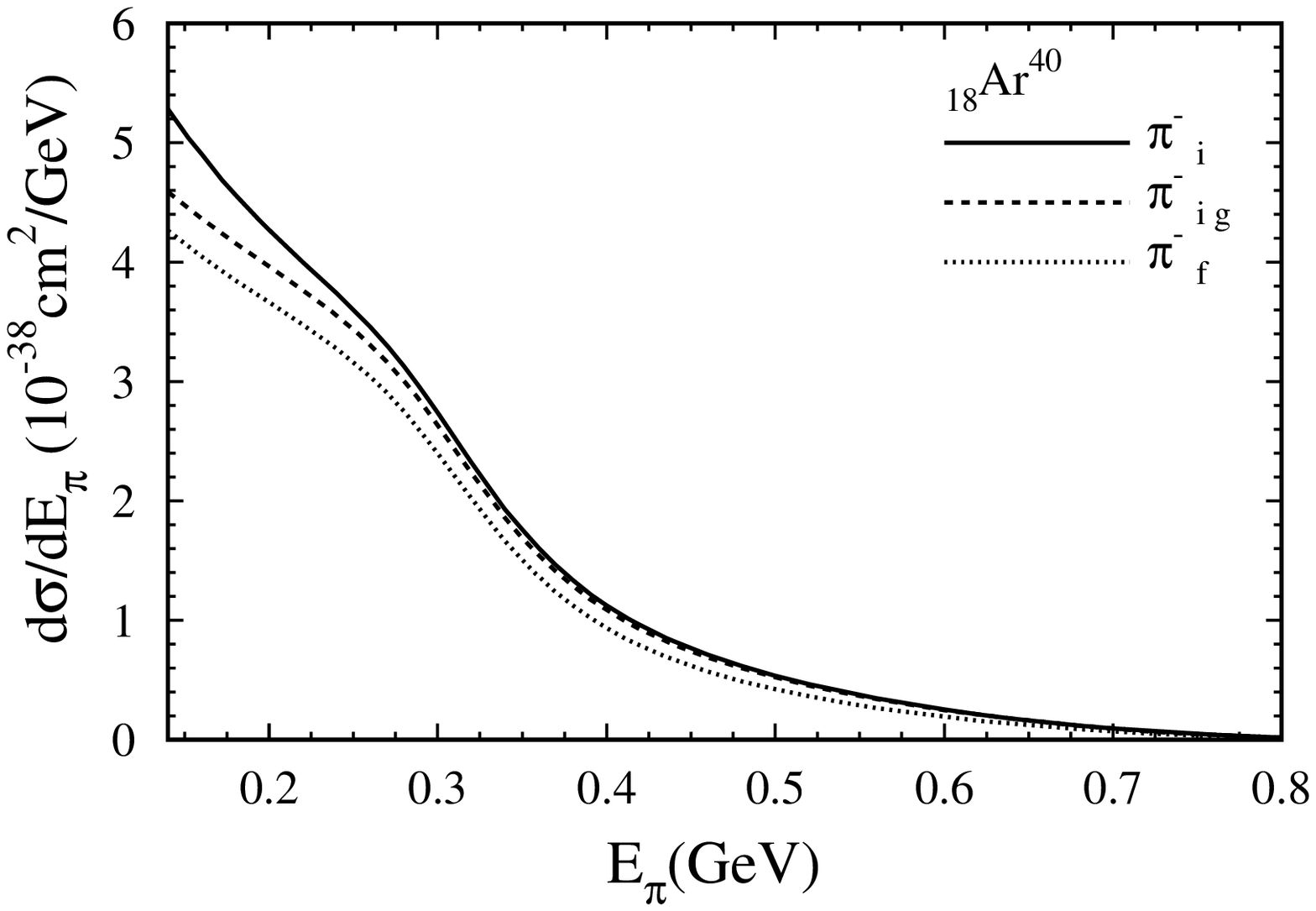,height=3.5in,angle=0}}
\caption{Pion energy distribution for negatively charged pions produced on argon targets. The solid, dashed and dotted lines represent respectively the pion energy distribution without any nuclear correction, including only the Pauli production factor $g$ and including all nuclear corrections.}
\end{figure}

%figure 14
\begin{figure}
\centerline{\psfig{figure=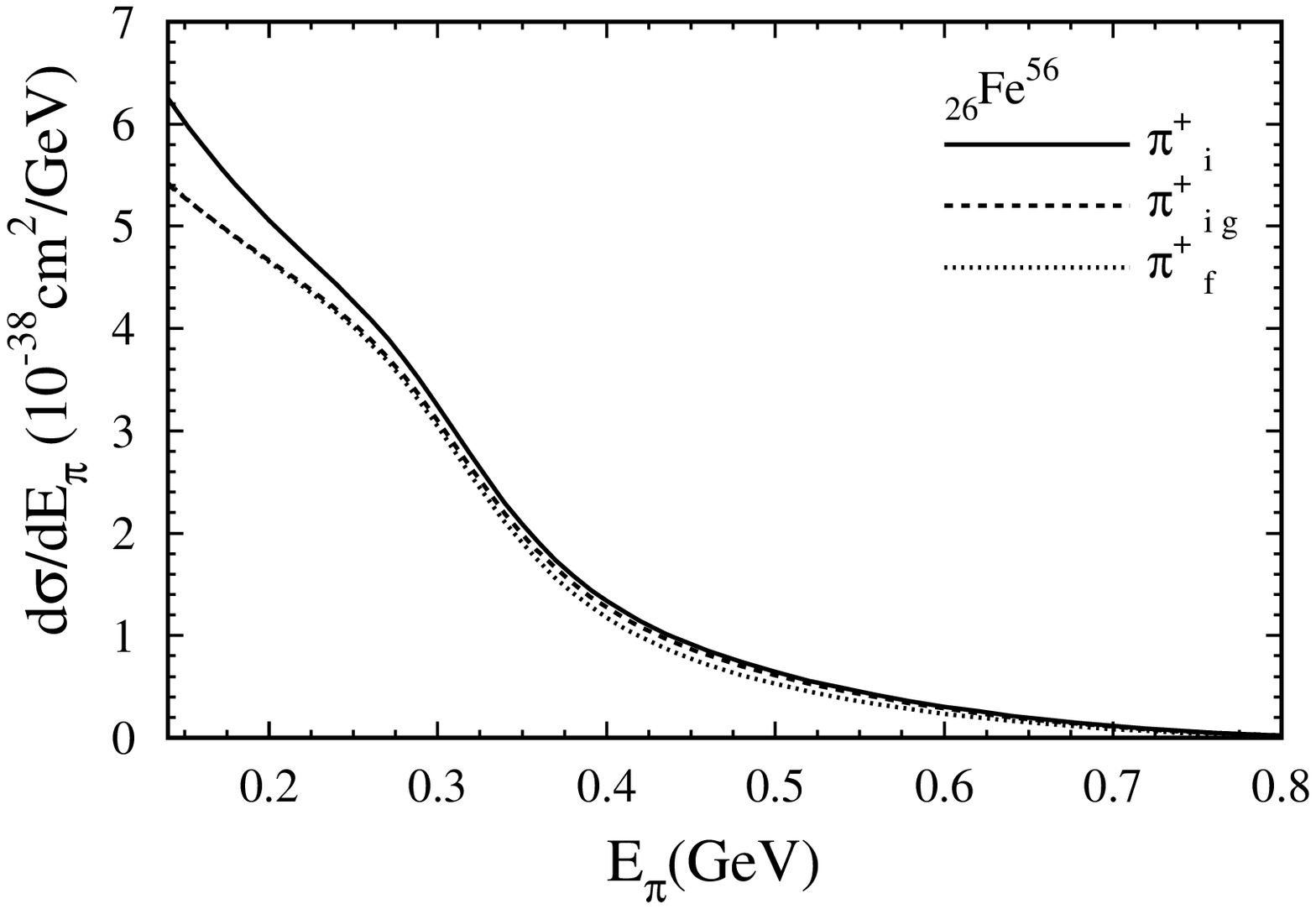,height=3.5in,angle=0}}
\caption{Pion energy distribution for positively charged pions produced on iron targets. The solid, dashed and dotted lines represent respectively the pion energy distribution without any nuclear correction, including only the Pauli production factor $g$ and including all nuclear corrections.}
\end{figure}

%figure 15
\begin{figure}
\centerline{\psfig{figure=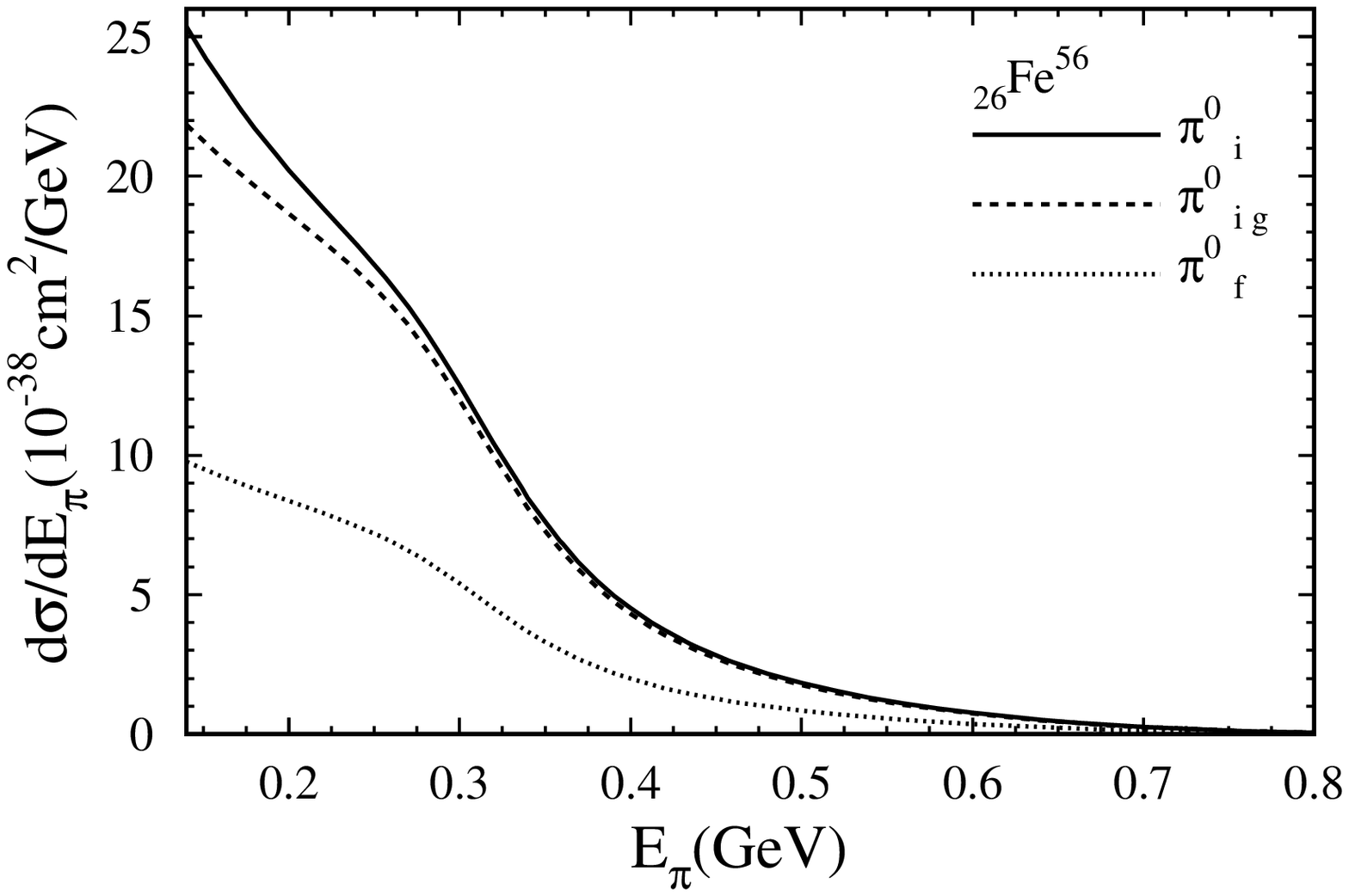,height=3.5in,angle=0}}
\caption{Pion energy distribution for neutrally charged pions produced on iron targets. The solid, dashed and dotted lines represent respectively the pion energy distribution without any nuclear correction, including only the Pauli production factor $g$ and including all nuclear corrections.}
\end{figure}

%figure 16
\begin{figure}
\centerline{\psfig{figure=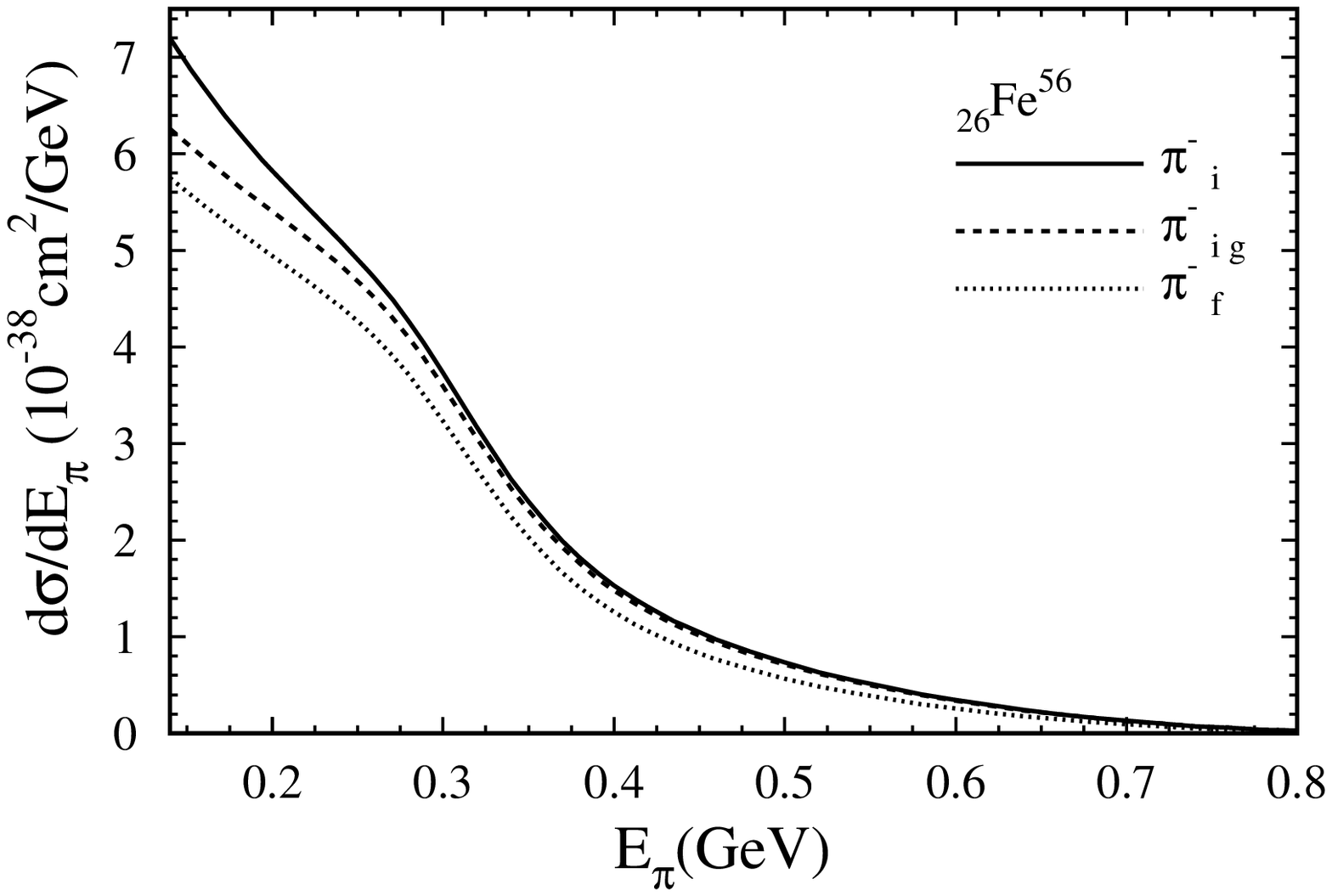,height=3.5in,angle=0}}
\caption{Pion energy distribution for negatively charged pions produced on iron targets. The solid, dashed and dotted lines represent respectively the pion energy distribution without any nuclear correction, including only the Pauli production factor $g$ and including all nuclear corrections.}
\end{figure}

\end{document}